\documentclass[cup6b]{cupbook}
\usepackage{graphicx}
\usepackage{natbib}
\title[Rome, Italy, 27--30 April 2009]
      {The coming of age of X-ray polarimetry}
\author{}
\date{}
\begin{document}
\pagenumbering{arabic}


\author[D. Lazzati]{Davide Lazzati \\ (Department of Physics, 
North Carolina State University,  \\ Riddick Hall, Box 8202, Raleigh, NC 27695-8202)}
\chapter{X-ray Polarization of Gamma-Ray Bursts}

\abstract{The degree and the temporal evolution of linear polarization
in the prompt and afterglow emission of gamma-ray bursts is a very
robust diagnostic of some key features of gamma-ray bursts jets and
their micro and macro physics. In this contribution, I review the
current status of the theory of polarized emission from GRB jets during
the prompt, optical flash, and afterglow emission. I compare the
theoretical predictions to the available observations and discuss the
future prospect from both the theoretical and observational
standpoints.}

\section{Introduction}

Gamma-Ray Bursts (GRBs) are the brightest explosions in the present day
Universe. Unfortunately, our understanding of their physics is still
incomplete, probably due to the fact that they are short lived,
point-like sources.

Polarization is a formidable tool to improve our understanding of GRB
jets: their geometry, magnetization, and radiation mechanism could in
principle be pinned down with a comprehensive and time-resolved analysis
of linear polarization. Observationally speaking, however, polarization
is not easy to measure. So far, only the optical afterglow has robust
polarization measurements \cite{covino99,greiner03} but the diverse
features and the sensitivity of the models to datail has made their
interpretation, at best, controversial.

In this review, I describe the theory underlying the production of
polarized radiation in GRBs in their three main phases. I will focus on
X-ray polarization but the discussion will be general, since the
frequency dependence of GRB polarization is very weak, especially at
frequencies where Faraday rotation is not relevant.

\section{Prompt emission}

The interest in the properties of linear polarization in the prompt
emission of GRBs increased dramatically with the claim that the prompt
emission of GRB~021206 had a linear polarization fraction of $\sim80\%$
\cite{coburn03}. Even though the claim was subsequently put into doubt
\cite{rutledge04,wigger04}, it generated a suite of models against which
any subsequent polarization observation will be compared. In this section we
present these models and compare them to observations.

\begin{figure} 
\centering 
\includegraphics[scale=.5]{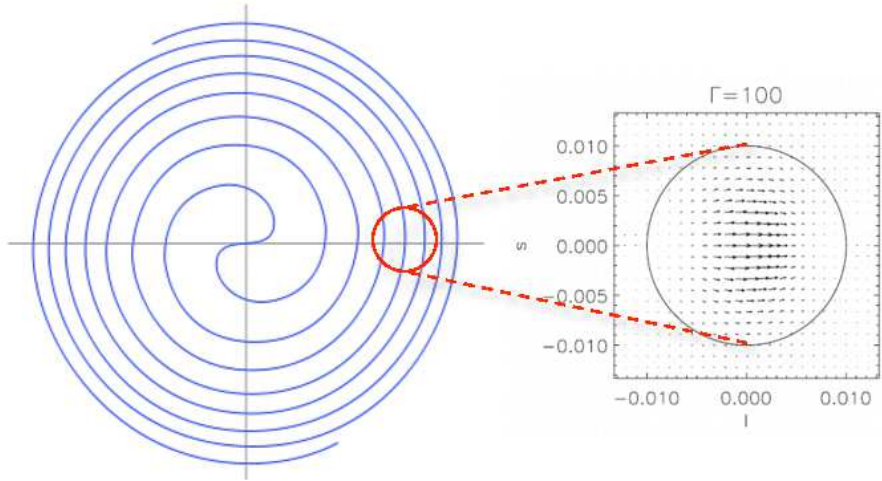}
\caption{Front view of the toroidal magnetic field in a fireball. The
circle highlights the fact that, due to relativistic aberration, only a
small fraction of the fireball is visible to the observer at infinite.
Additional relativistic aberration effects reduce the maximum
polarization, as shown in the right sub-panel \cite{lyutikov03}.}
\label{fig:1} 
\end{figure}

\begin{itemize}

\item {\bf Toroidal magnetic field model --- } Gamma-ray burst jets are
commonly believed to be produced by the effects of strong magnetization
(either of a neutron star\cite{bucciantini08} or of a massive accretion
disc onto a black hole\cite{macfadyen99}) combined with fast spinning.
In such conditions, sufficiently far from the jet engine, the magnetic
field is expected to be predominantly toroidal.

Synchrotron from a toroidal field configuration does not produce
polarized radiation in normal conditions. However, due to relativistic
aberration, only a small section of the whole toroidal structure
produces the radiation that is detected by an observer at infinity. As a
consequence, the radiation observed appears to come from a region of
uniform magnetic field and is maximally polarized (see
Fig.~\ref{fig:1}). Due to additional aberration effects, the
polarization angle direction is distorted in the edges of the visible
zone, slightly reducing the maximum detectable polarization
\cite{lyutikov03}.

Independently of details, this model predicts that almost all GRBs are
strongly polarized during their prompt emission. The polarization
position angle does not change with time, since the electric field
vector always points towards the pole of the field, i.e., the jet axis.
Only a very small fraction of bursts, those seen within an angle
$1/\Gamma$ from the jet axis, should display little to no polarization.

\item {\bf $1/\Gamma$ viewing angle effects ---} The main reason why
most models for GRB polarization predict small values of polarization
(few to ten per cent) is due to the fact that typically an observer
collects radiation from different regions with different polarization
orientations and the net signal is small. If, however, the fireball
configuration is such that only one emission zone is observed, high
polarization can be detected by the observer at infinity
\cite{waxman03}. Consider a fireball with an opening angle
$\theta_j\sim1/\Gamma$ observed at an angle $\theta_o=1/\Gamma$ from its
edge. Due to relativistic aberration, in the comoving frame the fireball
velocity and the line of sight are at a right angle. Both synchrotron
radiation from a planar magnetic field \cite{ghisellini99,
sari99,granot03} and bulk inverse Compton radiation \cite{lazzati04} are
maximally polarized in that configuration.

This model can in principle account for polarization up to 100 per cent.
Differently from the toroidal field model, only a small fraction of GRBs
should be polarized, due to the low probability for the particular
viewing configuration to be attained. As discussed for the toroidal
model the electric vector points towards the jet axis and the
polarization angle is therefore constant throughout the prompt emission
evolution.

\begin{figure} 
\centering 
\includegraphics[scale=.5]{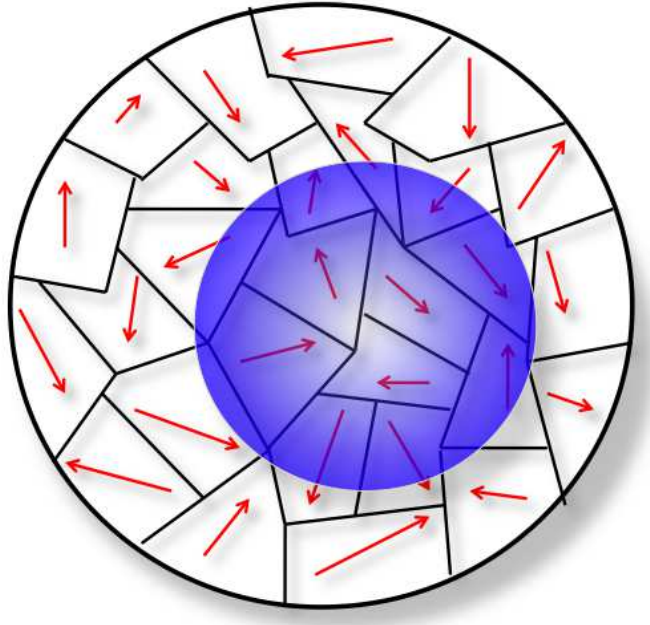}
\caption{Cartoon of the front view of a fireball with magnetic domains.
The arrows show the direction of the field in the domains. The shaded
circle emphasizes the fact that only some of the domains are visible by
the observer.}
\label{fig:2} 
\end{figure}

\item {\bf Magnetic domains --- } If the magnetic field generated by a
relativistic collisionless shock can reorganize into a uniform
configuration, the fireball surface would be covered with magnetic
patches, each with a different field orientation, but with a uniform
field within \cite{gruzinov99}. As a result of the speed of the field
re-organization and of relativistic aberration, the observer at infinity
sees radiation from approximately $N\sim100$ domains. The resulting net
polarization is therefore reduced by a factor $\sqrt{N}\sim10$. This
model predicts that all GRBs should be mildly polarized (in teh 10 per
cent range), with rapid fluctuations of the polarization angle. The
model cannot account for very high polarized fractions, as those
possibly observed in the prompt emission of GRBs.

\begin{figure} 
\centering 
\includegraphics[scale=.5]{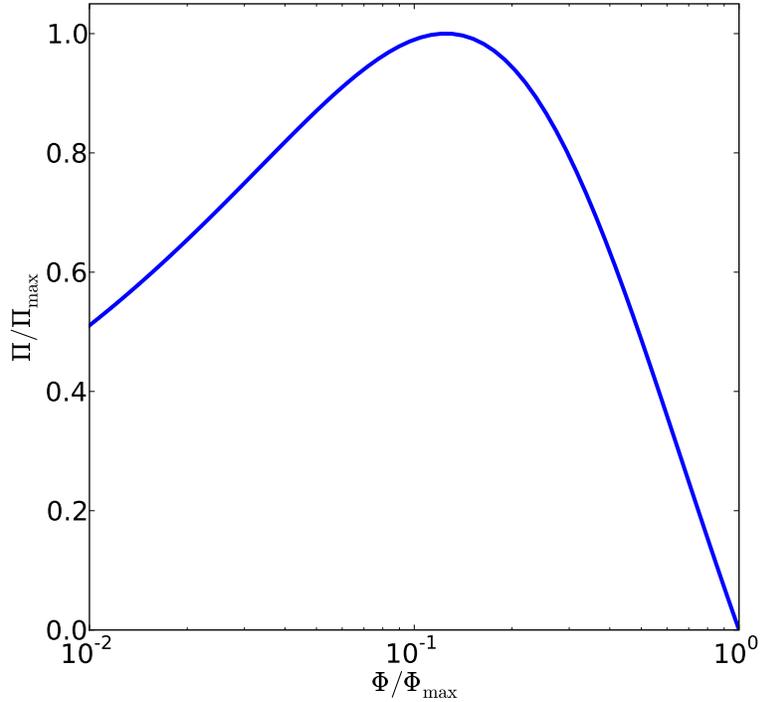}
\caption{Polarization vs. intensity for the radiation coming from a
fireball made by a large number of identical fragments with negligible
opening angle ($\theta_{\rm{jet}}\ll1/\Gamma$).}
\label{fig:3} 
\end{figure}

\item {\bf Fragmented fireballs --- } The main weakness of the
``$1/\Gamma$ effects" model is that it requires a very unlikely viewing
configuration. Such limitation would not be present if the fireball is
fragmented in shotguns \cite{heinz99}, cannonballs \cite{dado09}, or
mini-jets \cite{yamazaki06}. If we model a fireball as a series of
identical fragments, each producing radiation with the same efficiency
and moving at the same speed, polarization and intensity from a fragment
are strictly correlated \cite{lazzati09}. The brightest light observed 
comes from the fragment that is exactly pointing at the observer. Due to
cylindrical simmetry, the radiation is unpolarized. At the $1/\Gamma$
configuration, the radiation intensity is decreased by a factor
$\sim10$, and the polarization is maximum. For viewing angles
$\theta_o>1/\Gamma$, both the intensity and the polarization decrease
(see Fig.~\ref{fig:3}). Most bursts from fragmented fireballs are highly
polarized if a time-resolved analysis is performed, but they are weakly
polarized if the whole prompt emission is considered. That is because
the electric vector points towards each fragment, and so the position
angle fluctuates randomly from a pulse to the next.

\end{itemize}

\section{Afterglow}

\begin{figure} 
\centering 
\includegraphics[scale=.75]{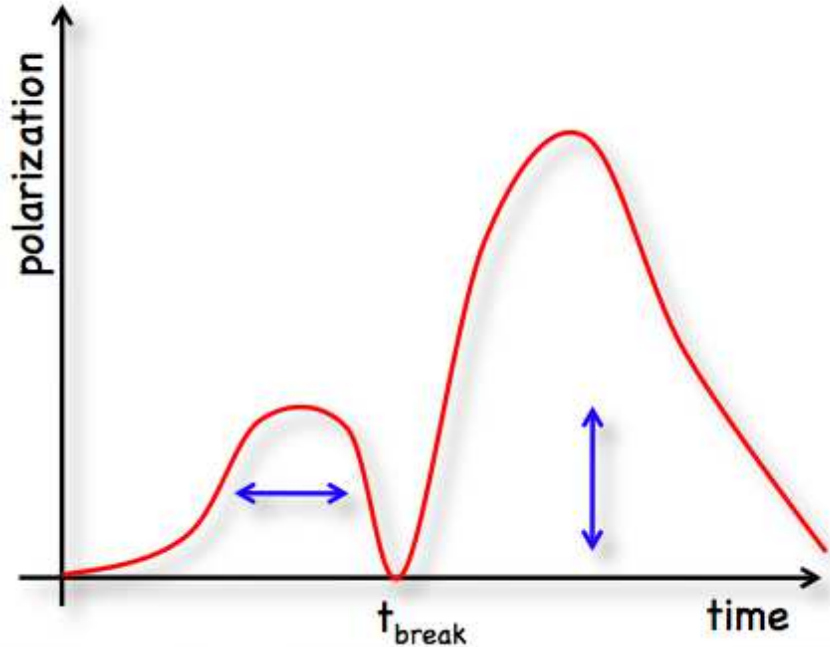}
\caption{Polarization of afterglow radiation from a uniform fireball.
the arrows indicate the direction of polarization.}
\label{fig:4} 
\end{figure}

The polarization of afterglow radiation has been observed with robust
results, but the comparison of observational data with models is
difficult. Afterglow radiation is known to be produced by synchrotron
from relativistic electrons gyrating into a shock-generated magnetic
field. Detailed calculations show that the polarization from a uniform
fireball is intimately connected to the evolution of the light curve
\cite{ghisellini99,sari99} and is very weakly dependent on the frequency
of photons (at least above optical frequencies, \cite{rossi04}).
Initially, the polarization is vanishingly small. At times before the
jet break, a small polarization of a few per cent is observed, with a
position angle perpendicular to the direction towards the jet axis. At a
time approximately coincident with the jet break time, the polarization
vanishes again. Subsequently, it reappears, rotated by 90 degrees,
reaches a maximum of $\sim10$ per cent and eventually vanishes again
(see Fig.~\ref{fig:4}).

The characteristic behavior of a 90 degrees rotation of the polarization
angle at a time roughly coincident with the jet break time is in
principle a formidable prediction and was actively looked for in
observations, with no success \cite{lazzati03,lazzati04b}. It was
subsequently realized that the polarization curve is very sensitive to
the brightness profile of the fireball, and that fireballs with a bright
core and less energetic wings produce a completely different
polarization curve, with maximum polarization around the break time and
a constant position angle \cite{rossi04}. Even more complicate is the
case of a fireball with bright spots randomly distributed on the
emitting surface. The polarization in that case would be virtually
unpredictable.

\section{Early afterglow}

The polarization of the prompt emission is in principle full of
important information to understand the physics of GRB jets. However,
polarization in the X-ray and $\gamma$-ray regimes is hard to observe.
Optical polarization is relatively easy to observe. However, models are
too sensitive to details and we haven't been able to obtain much robust
clues from optical polarization measurements. A potentially game changer
is polarization of the early optical afterglow, also known as the
optical flash. The optical flash is believed to be due to electrons in
the fireball energized by the reverse shock \cite{sari99b}. If that is
the case (see \cite{beloborodov02} for alternative models) the optical
flash should have the same polarization characteristic of the prompt
emission (and therefore carry a lot of insight) combined with the same
ease of observation of the afterglow polarization \cite{lazzati04b}.

\section{Discussion}

After discussing the polarization of the various stages of GRB emission
in detail, we here compare them with each other and with observations
and focus more on the X-ray aspects and future perspectives. Prompt
emission polarization is certainly the most appealing from the
theoretical point of view. Models are able to deliver univocal
interpretation for the various observational scenarios. The few
available observations are, however, inconclusive and contradictory.
Early observations claimed a high polarization for the overall burst
\cite{coburn03}. More recent observations find, instead, that the
polarization is indeed large, but the position angle varies from pulse
to pulse \cite{gotz09}. The observations are different and so too are
the implications. Constant position angle and high polarization point to
a toroidal magnetic field model, while variable angle is indicative of a
fragmented fireball scenario. The perspective of an early afterglow
polarization measurement is exciting, but the optical flash has no
emission in the X-rays, and its theoretical interpretation is still a
matter of open debate. While a positive measurement of large
polarization would be interesting, a no-polarization result, as the one
for GRB~060418 \cite{mundell07}, would be open to very many
interpretations. Afterglow observations are plagued by the model
sensitivity to details, with the notable exception of polarization of
the X-ray flares (Fan, this volume) which are supposed to be due to
engine activity and could therefore be polarized in the same way the
prompt emission is.

At the end of the day, what would be the best choice for an X-ray
polarimeter? Disregarding technical challenges, a theoretician would try
to observe prompt emission first, X-ray flashes second, and the
afterglow emission only as a last resort.

\end{document}